%% file: ms.tex
\documentclass[12pt,preprint]{aastex}

\shorttitle{Study of Post Flare Loops} \shortauthors{Li et al.}

\begin{document}

\title{On the Brightening Propagation of Post-Flare Loops Observed by TRACE}

\author{Leping Li and Jun Zhang }

\affil{National Astronomical Observatories, Chinese Academy of
Sciences, Beijing 100012, China; lepingli;zjun@ourstar.bao.ac.cn}

\begin{abstract}

Examining flare data observed by TRACE satellite from May 1998 to
December 2006, we choose 190 (151 M-class and 39 X-class) flare
events which display post-flare loops (PFLs), observed by 171
\AA~and 195 \AA~wavelengths. 124 of the 190 events exhibit flare
ribbons (FRs), observed by 1600 \AA~images. We investigate the
propagation of the brightening of these PFLs along the neutral
lines and the separation of the FRs perpendicular to the neutral
lines. Observations indicate that the footpoints of the initial
brightening PFLs are always associated with the change of the
photospheric magnetic fields. In most of the cases, the length of
the FRs ranges from 20 Mm to 170 Mm. The propagating duration of
the brightening is from 10 minutes to 60 minutes, and from 10
minutes to 70 minutes for the separating duration of the FRs. The
velocities of the propagation and the separation range from 3 km
s$^{-1}$ to 39 km s$^{-1}$ and 3 km s$^{-1}$ to 15 km s$^{-1}$,
respectively. Both of the propagating velocities and the
separating velocities are associated with the flare strength and
the length of the FRs. It appears that the propagation and the
separation are dynamically coupled, that is the greater the
propagating velocity is, the faster the separation is.
Furthermore, a greater propagating velocity corresponds to a
greater deceleration (or acceleration). These PFLs display three
types of propagating patterns. Type I propagation, which possesses
about half of all the events, is that the brightening begins at
the middle part of a set of PFLs, and propagates bi-directionally
towards its both ends. Type II, possessing 30$\%$, is that the
brightening firstly appears at one end of a set of PFLs, then
propagates to the other end. The remnant belongs to Type III
propagation which displays that the initial brightening takes
place at two (or more than two) positions on two (or more than
two) sets of PFLs, and each brightening propagates
bi-directionally along the neutral line. These three types of
propagating patterns can be explained by a three-dimensional
magnetic reconnection model.
\end{abstract}


\section{INTRODUCTION}

Flares are one of the most spectacular phenomena in solar physics.
They are sudden brightening in the solar atmosphere, and consist
of a number of components including loops, ribbons, arches, remote
patches, surges, erupting filaments, and other expanding coronal
features \citep{mar89}. They have been studied morphologically
from direct images \citep[e.g.][]{kru00,fle01,kun01} and
spectroscopically from spectrograms
\citep[e.g.][]{moo76,cow73,act85,cul97,gri05b} at different
wavelength regions. Like most dynamic phenomena on the solar
surface, the occurrence of solar flares is closely related to the
presence and evolution of solar magnetic fields, especially the
complicated, non-potential magnetic configuration
\citep{rus72,pat81,moo84}. Flares can be caused by rotating
sunspot \citep{bro03,tia06,zha07}, magnetic flux emergence
\citep{ish98,wan04,li07}, magnetic flux cancellation
\citep{liv89,wan92,wan93,zha01a,zha01b,zha02}, magnetic shear
\citep{kus04,wan06,ji06,su07}, and so on.

The chromospheric flare \citep[e.g.][]{fan00,fal02,che06,hud07} is
easier to be observed, especially with the help of an H$\alpha$
filter. Larger flares often occur right after the sudden
disappearance of a filament
\citep{kup81,din03,ste05,jia06a,jia06b,chi07}. In this case, the
flare generally has the form of two flare ribbons (FRs) which lie
on both sides of the location of the former filament. During flare
decay phase, the FRs move apart. The separation of these FRs has
been used to estimate the electric field in the reconnecting
current sheet \citep[e.g.][]{qiu02,asa04b} and also the coronal
magnetic field strength and the reconnection rate
\citep[e.g.][]{iso02b,iso05}. Many authors have observed the
parallel and antiparallel movements of the FRs along the arcade
\citep{hoy81,tak83,fle02,liu04,qiu04,sia04}.

Accompanying the FRs is a system of post-flare loops (PFLs) which
initially appears at low altitude and then moves upward into the
corona in consort with the motion of the ribbons \citep{moo80}. A
classical description of the PFLs, as seen in H$\alpha$ images,
was first given by \citet{bru64} who noted that the ribbons
essentially lie at the footpoints of the loop system, which forms
a series of arcades \citep{lin03}. These arcades are also
frequently seen in the X-ray images recorded by Soft X-ray
Telescope (SXT) \citep{tsu91} aboard Yohkoh \citep{oga91}, in the
extreme ultraviolet (EUV) images from EIT
\citep[Extrmeme-ultraviolet Imaging Telescope,][]{del95} aboard
the Solar and Heliospheric Observatory (SOHO) \citep{dim95}, and
also in the EUV images from Transition Region and Coronal Explorer
(TRACE) \citep{han99}. The formation of transient large-scale PFLs
or post-eruptive arcades (PEAs) has been widely studied
\citep[e.g.][]{car64,stu66,hir74,kop76,car83,web87,sve97,tri04,tri05,tri06b}.
\citet{hud98}, \citet{ste00} and \citet{tri06b} studied a set of
individual events about the relationship between PFLs and Coronal
Mass Ejections (CMEs). The propagation of the loop formations
along the neutral line, together with the separation of the FRs
perpendicular to the neutral line was reported by \citet{iso02a}.
\citet{gri05a} found a RHESSI observation showing that the hard
X-ray (HXR) sources do not show the separation from the neutral
line, but instead they move along the neutral line \citep[see
also][]{gof07}. \citet{tri06a} investigated the relationship
between the brightening propagation of the PEAs and the erupting
filament/prominence. They reported two types of propagation of the
brightening, and the propagating direction was consistent with the
erupting direction of the filaments.

Due to the extremely low density and high temperature of the
corona, measurements of the magnetic field are restricted to lower
layers of the solar atmosphere. For this reason, extrapolation
techniques, which attempt to reconstruct the coronal field from
measured boundary values in the photosphere (or low chromosphere),
are the prime tool for quantitative investigations of the coronal
magnetic field \citep{val05}. Some authors used the methods of
extrapolation to study the evolution of flare loops in active
region \citep[e.g.][]{yan95,wan01,wie05,ama06,zha08}.

Magnetic reconnection of solar coronal loops is considered to be
the main process that causes solar flares and possibly coronal
heating. It is widely believed to be a mechanism of magnetic
energy release \citep[see reviews by][]{shi99,mar03}, and plays an
important role in various explosive phenomena in astrophysical,
space, and laboratory plasmas \citep{bis93,taj97,pri00}. The
evidence of magnetic reconnection found by space observations
includes the cusp-shaped PFLs \citep{tsu92}, the loop-top hard
X-ray source \citep{mas94}, the reconnection inflow
\citep{yok01,lin05}, downflows above PFLs
\citep{mck99,inn03,asa04a}, plasmoid ejections
\citep{shi95,ohy97,ohy98}, etc. Two-dimensional
\citep[e.g.][]{dey71,win91,hu95,for95,yok98,zha06} and
three-dimensional simulations
\citep[e.g.][]{wu92,mag99,miy04,aul06,bir06} are widely used to
study the magnetic reconnection in process of flares and CMEs.

The magnetic reconnection model proposed by \citet{car64},
\citet{stu66}, \citet{hir74} and \citet{kop76} (the CSHKP model)
suggests that magnetic field lines successively reconnect in the
corona. This model explains several well-known features of solar
flares, such as the growth of flares loops with a cusp-shaped
structure and the formation of the H$\alpha$ two-ribbon structures
at their footpoints. In recent decades, this model has been
further extended
\citep[e.g.][]{pne81,pri90,pri00,pri02,moo92,moo01,shi99,yok01,lin00,lin04}.

In this paper, we mainly study the dynamic evolution of a larger
sample of flare events, including the propagation of the PFLs
along the neutral lines and the separation of the FRs away from
the neutral lines. This investigation will provide some
information of 3-dimensional magnetic reconnection in the process
of flare. The criteria for the data selection and the methods of
the data analysis are described in section 2. In section 3, we
summarize our statistical results. Conclusions and brief
discussion are shown in section 4.

\section{DATA AND OBSERVATIONS}

We checked all the 75 X-class and 509 M-class flare
events\footnote{http://hea-www.harvard.edu/trace/flare\_catalog/index.html}
observed by TRACE from May 1998 to December 2006. 39 X-class and
151 M-class flare events which display PFLs are chosen as our
sample (see Table 1). The TRACE mission explores the dynamics and
evolution of the solar atmosphere from the photosphere to the
corona with high spatial and temporal resolution~\citep{han99}. It
observes the photosphere (white-light, WL), the transition region
(1216, 1550, and 1600 \AA) and the 1-2 MK corona (171, 195, and
284 \AA). In this work, we mainly use the 171 and 195 \AA~images
to study the propagation of the brightening of the PFLs, and the
1600 \AA~observations to the separation of the FRs. We also use
TRACE WL observations and magnetic field observations of SOHO/MDI
\citep{sch95} to investigate the variation of the source region of
these flare events.

\subsection{Methods}

In order to study the dynamic evolution of the larger sample of
flare events quantificationally, we determine a set of parameters
to describe the evolution of the PFLs and the FRs for each event
in our sample. These parameters include the propagating duration
(\emph{PD}), propagating velocity (\emph{V$_{p}$}) and
acceleration (\emph{a$_{p}$}) of the PFLs, the separating duration
(\emph{SD}) and separating velocity (\emph{V$_{s}$}) of the FRs,
the length (\emph{Len}) of the FRs at the maximum of the flare,
the flare class (\emph{FC}) and the flare duration (\emph{FD}).
Because some events in the sample are not completely observed, we
can not get all the parameters simultaneously for all the cases.
Among the 190 flare events, 183 display clear evolution of the
PFLs observed by TRACE 171 and 195 \AA~images, so we can measure
the three parameters \emph{V$_{p}$}, \emph{a$_{p}$} and \emph{PD}.
TRACE 1600 \AA~observations can clearly detect the FRs. In our
sample, 124 events are also observed by TRACE 1600 \AA~wavelength,
so we measure the parameter \emph{Len} for these events. 101 of
the 124 events show clear kinetics of the FRs, we measure
\emph{V$_{s}$} and \emph{SD} for these cases. Table 2 lists the
measured events of these parameters.

In order to illustrate how to measure these parameters, we show a
flare event observed on the disk at the heliographic position S05
W54 on 2001 March 20 in Fig. 1. Figure 1\emph{a} shows the initial
brightening of PFLs at 02:41 UT. \emph{L$_{21}$} and
\emph{L$_{22}$} in Fig. 1\emph{b} represent the distance of the
propagation of the PFLs towards southeast along different FRs from
02:41 UT to 03:16 UT, as well as \emph{L$_{11}$} and
\emph{L$_{12}$}, towards northwest. The total value of these four
distance represents the distance of the propagation
(\emph{L$_{p}$}) of the PFLs from 02:41 UT to 03:16 UT:
\begin{equation}
L_{p}=L_{11}+L_{12}+L_{21}+L_{22}~.
\end{equation}
\emph{L$_{31}$} and \emph{L$_{32}$} in Fig. 1\emph{c} show the
distance of the propagation of the PFLs toward southeast from
02:41 UT to 03:52 UT, as well as \emph{L$_{41}$} and
\emph{L$_{42}$}, towards northwest. The distance of the
propagation of PFLs from 02:41 UT to 03:52 UT is:
\begin{equation}
L_{p}=L_{31}+L_{32}+L_{41}+L_{42}~.
\end{equation}
By using the method shown above, we obtain a series of propagating
distance of the PFLs with time, and a series of propagating
velocities is derived. Both the initial propagating time
(\emph{t$_{pi}$}) and the end propagating time (\emph{t$_{pe}$})
of the brightening of the PFLs are determined from the series of
velocities. We define \emph{t$_{pe}$} to be the time when the
propagation velocity drops to 1/e of the peak value. The parameter
\emph{PD} of the PFLs is the duration of \emph{t$_{pi}$} and
\emph{t$_{pe}$}. The distance versus time plot for the propagation
of the PFLs between \emph{t$_{pi}$} and \emph{t$_{pe}$} for the
event showed in Fig. 1 is indicated in Fig. 2. We use a linear
polynomial fit to the data points to get the parameters
\emph{V$_{p}$}, and second order polynomial fit for
\emph{a$_{p}$}. Similar to the determination of \emph{t$_{pi}$}
and \emph{t$_{pe}$}, the initial separating time (\emph{t$_{si}$})
and the end separating time (\emph{t$_{se}$}) of the FRs are
obtained from 1600 \AA~observations. The parameter \emph{SD} is
the duration from \emph{t$_{si}$} to \emph{t$_{se}$}. Figures
1\emph{d}-1\emph{f} are series of 1600 \AA~images observed by
TRACE. \emph{L$_{51}$} and \emph{L$_{52}$} in Fig. 1\emph{f}
represent the distance of the separation of the FRs from 02:25 to
03:27 UT towards northeast and southwest, respectively. The
distance of the separation (\emph{L$_{s}$}) of the FRs is
\begin{equation}
L_{s}=L_{51}+L_{52}~.
\end{equation}
We got the parameter \emph{V$_{s}$} using the similar method as to
get \emph{V$_{p}$}. The length of the FRs is measured from 1600
\AA~image (see Fig. 1\emph{e}) while the FRs are well developed.
\emph{L$_{61}$} and \emph{L$_{62}$} represent the length of two
FRs, respectively. We get \emph{Len} by using
\begin{equation}
Len=\frac{L_{61}+L_{62}}{2}.
\end{equation}
The projection effects are corrected using trigonometric function
when we got the parameters \emph{V$_{p}$}, \emph{a$_{p}$},
\emph{V$_{s}$} and \emph{Len}.

The parameter \emph{FC}, the peak X-ray flux of the associated
flare, is recorded by GOES-10 soft X-ray 1-8 \AA~flux intensity,
and \emph{FD}, the duration between the beginning time and the
ending time of the flares, by Solar-Geophysical Data. While
studying the evolution of the PFLs of these events, we find that
the PFLs display three types of propagating patterns. Type I is
that the brightening begins at the middle part of a set of PFLs,
and then propagates bi-directionally towards its both ends. Type
II shows that the brightening firstly appears in one end of a set
of PFLs, then propagates to the other end. There are two (or more
than two) initial brightening involved in Type III propagation,
each brightening takes place in the middle of a set of PFLs, then
propagates bi-directionally. The following are three examples of
flare events which are displayed to describe the three types of
propagation.




\subsection{M 9.8 Flare on 2005 September 17}

Type I propagation is characterized by an M 9.8 flare event
occurred at the heliographic position S11 W51 on 2005 September
17. We show the time sequence of 171 \AA~images in the top panels
of Fig. 3. The brightening of the PFLs is firstly seen at 06:04
UT, and the outer edges of the PFLs are outlined as dotted curves
in Fig. 3\emph{a}. The propagation of the PFLs ended at 06:24 UT
(see Figure 3\emph{c}). From these panels, we notice that the
initial brightening of this event occurs at the middle part of a
set of PFLs, then propagates bi-directionally towards its both
ends (see white arrows in Fig. 3\emph{c}).

We examine the evolution of photospheric magnetic field and the
change of continuum intensity by using the observations of
SOHO/MDI and TRACE WL, respectively. Figure 3\emph{d} is the
magnetogram before the flare, as well as Fig. 3\emph{e}, after the
flare. Figure 3\emph{f} displays the difference image of these two
magnetograms. The regions in white brackets show the change of
photospheric magnetic field, and the variation of unsigned
magnetic flux is about 3$\times$10$^{20}$ Mx. The TRACE WL images
before and after the flare are presented in Figs. 3\emph{g} and
3\emph{h}, respectively. We show their difference image in Fig.
3\emph{i}, and find the change of the continuum intensity (denoted
by white brackets). Comparing Figs. 3\emph{f} and 3\emph{i} with
Figs. 3\emph{a} and 3\emph{c}, we note that the regions showing
obvious changes of photospheric magnetic field and continuum
intensity are associated with the initial brightening of the PFLs
(see the brackets in Figs. 3\emph{a} and 3\emph{c}). In other
words, the initial brightening is related to the magnetic
variation on the photosphere.

\subsection{M 2.0 Flare on 1999 January 18}

The M 2.0 flare event on 1999 January 18, which occurred at the
heliographic position N19 E03, is used to describe Type II
propagation. Figures 4\emph{a} and 4\emph{c} are the TRACE 171
\AA~images showing the beginning and the ending propagation of the
brightening of the PFLs. Figure 4\emph{b} is an image during the
propagation. It appears that the initial brightening occurs at the
southern end of a set of PFLs, and then propagates towards
northeastern end (see the white arrow in Fig. 4\emph{c}).

As this event was not observed by SOHO/MDI, we use TRACE WL images
to study the change of the continuum intensity (the bottom panels
of Fig. 4) which are relevant to this event. Figures 4\emph{d} and
4\emph{e} are the observations before and after the flare, and
Fig. 4\emph{f} is their difference image. The regions inside the
square brackets in Fig. 4\emph{a} show the footpoints of the
initial brightening of the PFLs. We overlay these square brackets
on Fig. 4\emph{f}, and find that the initial brightening of the
PFLs is associated with the change of the continuum intensity.

\subsection{M 1.5 Flare on 2002 October 25}

The M 1.5 flare event on 2002 October 25 occurred at the
heliographic position N28 W11. This event is employed to display
Type III propagation. Figure 5\emph{a} shows the initial
brightening of the PFLs in the southwest of the FRs, with dotted
curves outlining the outer edge of the PFLs. Then this brightening
propagates towards northeast and southwest (see the black arrows
showed in Fig. 5\emph{c}). Figure 5\emph{b} presents the
brightening of another set of PFLs (marked by the dashed line).
This brightening also propagates towards northeast and southwest
(see the white arrows in Fig. 5\emph{c}). From these panels, we
notice that the PFLs are consist of two sets of independent
brightening loops in different places in the active region, then
each brightening propagates bi-directionally towards its both
sides.

Figures 5\emph{d} and 5\emph{e} are longitudinal magnetograms
before and after the flare observed by SOHO/MDI, respectively.
Figure 5\emph{f} shows their difference signal. We mark the
regions of the PFL footpoints as brackets in Figs.
5\emph{a}-5\emph{b} and overlay these regions on the difference
magnetogram (see Fig. 5\emph{f}). It indicates that the regions of
the initial brightening are associated with the change of the
photospheric magnetic field, as presented in section 2.2.

\section{STATISTICAL RESULTS}

In this work, eight parameters (\emph{PD}, \emph{V$_{p}$} and
\emph{a$_{p}$} of PFLs, \emph{SD}, \emph{V$_{s}$} and \emph{Len}
of FRs, \emph{FC} and \emph{FD}) are considered to characterize
the kinematics of the PFLs and the FRs, and listed in Table 2.

By examining the TRACE observations of all the flare events, we
have determined \emph{PD} of 183 events and \emph{SD} of 101
events. Figure 6\emph{a} shows the distribution of \emph{PD}
(solid lines) and \emph{SD} (dotted lines). \emph{PD} ranges from
10 to 60 minutes in nearly 90$\%$ of the cases, with the peak of
the distribution lying close to 25 minutes. The average value of
all the \emph{PD}s is 33 minutes. Similarly, almost 90$\%$ of the
\emph{SD}s range from 10 to 70 minutes, with average duration of
38 minutes. Besides \emph{PD} and \emph{SD}, we have also
exhibited the distribution of \emph{V$_{p}$} (solid lines) and
\emph{V$_{s}$} (dotted lines) in Fig. 6\emph{b}. Most of the
\emph{V$_{p}$}s range from 3 to 33 km s$^{-1}$, and the average
velocity is 17.9 km s$^{-1}$. It indicates that 76$\%$ of the
\emph{V$_{s}$}s range from 3 to 15 km s$^{-1}$, with the average
value of 7.2 km s$^{-1}$.

In order to explore the physical connection of these parameters,
we study the one-to-one correspondence among them. It appears that
\emph{FD} is associated with \emph{PD} and \emph{SD}, and the
longer the \emph{FD} is, the longer the \emph{PD} and the
\emph{SD} are, as displayed in Fig. 7. The \emph{Len} of the FRs,
which ranges mainly from 20 to 170 Mm and may represent the length
of the current sheet along the neutral line, is associated with
\emph{FC} and weakly associated with \emph{FD} (see Fig. 8). It
looks like that more powerful flares correspond to longer FRs.
\emph{V$_{p}$} and \emph{V$_{s}$} are two important parameters, as
they represents the kinetics of the propagation of the PFLs and
the separation of the FRs. Figure 9 shows the correspondence of
\emph{V$_{p}$} and \emph{V$_{s}$} with some relevant parameters.
Generally, both of \emph{V$_{p}$} and \emph{V$_{s}$} increase as
the flare becomes more powerful (see Figs. 9\emph{a} and
9\emph{c}). Figures 9\emph{b} and 9\emph{d} show the relationship
between the velocity and \emph{Len} of the FRs, separately for
\emph{V$_{p}$} (Fig. 9\emph{b}), and \emph{V$_{s}$} (Fig.
9\emph{d}). It displays that both \emph{V$_{p}$} and
\emph{V$_{s}$} increase from several km s$^{-1}$ to tens of km
s$^{-1}$, as \emph{Len} increases from tens of Mm to hundreds of
Mm.

The separation of the FRs, considered to be the representation of
continuing magnetic reconnection and upwards moving reconnection
sites, has been well studied. The propagation of the PFLs, which
may reflect the signature of successive magnetic reconnection
along the neutral line, is rarely taken into account. In our
statistical results, we notice that \emph{V$_{s}$} is associated
with \emph{V$_{p}$} (see Fig. 10\emph{a}), that is \emph{V$_{s}$}
increases as \emph{V$_{p}$} increases. Besides the velocity,
\emph{a$_{p}$} is also an important parameter to study the
evolution of the PFLs. Figure 10\emph{b} shows the relationship
between \emph{a$_{p}$} and \emph{V$_{p}$}. The diamonds represent
the positive accelerations of the brightening propagation, as well
as the asterisks, decelerations. Among these 183 events, 135 are
decelerated, possessing 74$\%$, and 48 accelerated, occupying
26$\%$. There is also a trend that the greater the \emph{V$_{p}$}
is, the larger the deceleration (or acceleration) is.

In this study, we classify the propagation of the PFLs into three
types. The information of these three types of propagation is
showed in Table 3. It appears that the events displaying Type I
propagation are almost half of all the events.

\section{CONCLUSIONS AND DISCUSSION}

We have studied the evolution of the PFLs and the FRs of the 190
flare events, and obtained the following results.

1. Both of \emph{PD} and \emph{SD} are associated with \emph{FD}.
The longer the \emph{FD} is, the longer the \emph{PD} and the
\emph{SD} are. The length of the FRs mainly ranges from 20 to 170
Mm. It is associated with \emph{FC}, but weakly associated with
\emph{FD}. It increases as \emph{FC} (\emph{FD}) increases.

2. \emph{V$_{p}$} ranges mainly from 3 km s$^{-1}$ to 33 km
s$^{-1}$, and \emph{V$_{s}$}, from 3 km s$^{-1}$ to 15 km s$^{-1}$
in 76$\%$ of the cases. \emph{V$_{p}$} and \emph{V$_{s}$} are
dynamically coupled, \emph{V$_{p}$} increases as \emph{V$_{s}$}
increases. Both of \emph{V$_{p}$} and \emph{V$_{s}$} are
positively correlated with \emph{FC} and \emph{Len} of the FRs.
The brightening of the PFLs do not evenly propagate, 74$\%$ of the
PFL events are decelerated, and 26$\%$, accelerated.

3. There are three types of propagation of the PFLs. Type I,
possessing 49.5$\%$ of all the events, is that the brightening
begins at the middle part of a set of PFLs, and then propagates
bi-directionally towards its both ends. Type II, in possession of
30$\%$, displays that the brightening firstly appears in one end
of a set of PFLs, then propagates to the other end. Type III,
occupying 20.5$\%$, shows that the initial brightening takes place
at two (or more than two) positions on two (or more than two) sets
of PFLs, and each brightening propagates bi-directionally.

The main error source of our measurement of the velocities
(\emph{V$_{p}$} and \emph{V$_{s}$}) is due to the uncertainty of
the outer edges. In our study, two pixels error has been used. As
the average \emph{PD} (\emph{SD}) is 33 (38) minutes, the error of
\emph{V$_{p}$} (\emph{V$_{s}$}) is 0.4 (0.3) km s$^{-1}$, which is
much less than the velocities showed in Table 1. So the values of
these velocities are reliable.

Several authors have paid attention to the propagation of the
PFLs. \citet{iso02a} reported that \emph{V$_{p}$} is about 3-30 km
s$^{-1}$ by using the data from Yohkoh/SXT \citep{tsu91}, and
50-150 km s$^{-1}$ in \citet{gri05a} from RHESSI observations.
Based on the SOHO/EIT observations, \citet{tri06a} studied several
flare events relevant to long filament eruptions. They presented
that \emph{V$_{p}$} of the PEAs ranges from 20 to 111 km s$^{-1}$.
In this paper, we show that \emph{V$_{p}$} is mainly from 3 to 33
km s$^{-1}$, which is consistent with \citet{iso02a}, but somewhat
smaller than that of \citet{tri06a}. The velocity difference
between ours and Tripathi et al. may result from two aspects. The
first aspect is the sample. Our sample contains hundreds of cases.
The second is the datum source. We employ the TRACE data which
have higher spatial (1$\arcsec$) and temporal (1 minute)
resolution. \emph{V$_{s}$} of the FRs have been intensively
studied, but the value are varied distinctly, e.g. 20-100 km
s$^{-1}$ in \citet{qiu02}, 0-85 km s$^{-1}$ in
\citet{asa04b,asa06} and 20-70 km s$^{-1}$ in \citet{tem07}. By
employing the TRACE 1600 \AA~observations, \citet{iso05} obtained
\emph{V$_{s}$} of 6.7-12 km s$^{-1}$ near the impulsive phase of
the flares. Our results about \emph{V$_{s}$} (3-15 km s$^{-1}$)
are well in agreement with that of \citet{iso05}.

It is basically accepted that the separation of the FRs and the
propagation of the PFLs can be considered the signature of the
successive magnetic reconnection along different directions.
\emph{SD} represents the duration of the magnetic reconnection
where the reconnection points move upward, and \emph{PD}, along
the neutral lines. Both \emph{SD} and \emph{FD} determine the
duration of flares. That is the reason why both \emph{SD} and
\emph{PD} are associated with \emph{FD} (see Figs.
7\emph{a}-\emph{b}). Lots of researchers have employed
\emph{V$_{s}$} to estimate the reconnection rate, they derived
that the reconnection rate is very sensitive to \emph{V$_{s}$}
\citep[e.g.][]{iso02b,iso05}. Investigation of the propagation of
the PFLs and the separation of the FRs simultaneously will provide
us with 3-dimensional picture of magnetic reconnection in the
process of flares. In our study, we find that \emph{V$_{p}$} of
the PFLs is associated with \emph{V$_{s}$} of the FRs (see Fig.
10\emph{a}). This indicates that the separation of the FRs and the
propagation of the PFLs are dynamical coupled. Our results show
that both \emph{V$_{s}$} of the FRs and \emph{V$_{p}$} of the PFLs
are associated with \emph{FC} (see Figs. 9\emph{a} and 9\emph{c}).
As \emph{FC} represents the maximum energy release rate of the
flares, this implys that \emph{V$_{s}$} and \emph{V$_{p}$} may
also reflect the magnetic reconnection rate. \emph{Len} of the FRs
may be the length of the current sheet along the neutral line.
Both \emph{V$_{s}$} of the FRs and \emph{V$_{p}$} of the PFLs are
associated with \emph{Len} of the FRs (see Figs. 9\emph{b} and
9\emph{d}). This may result from the magnetic configuration of the
source region of the flare. \citet{tri06a} have suggested that the
longer the FRs is, the less magnetically complex the flare region
is. We speculate that the brightening of the PFLs propagates
easily in a simple magnetic configuration, and so do the FRs
separate. All the parameters and the relationships between these
parameters we discussed here will help us to better understand
3-dimensional magnetic reconnection, and provide an observational
character for theoretical study and simulation of 3-dimensional
magnetic reconnection.

By investigating 17 events, \citet{tri06a} found two types of
propagation of the brightening of the PEAs associated with
asymmetric eruptions and symmetric eruptions of filaments,
respectively. We have studied a much larger sample with higher
spatial and temporal resolution observations, and found three
types of propagation of the PFLs. Type I and Type II propagation
are similar to those two types of propagations reported by
\citet{tri06a}. Type III propagation indicates that sometimes two
or more than two sets of PFLs are involved in a flare event. All
the three types of the propagation of the PFLs may be explained by
schematic diagrams shown in Fig. 11 \citep[see
also][]{shi05,tri06a}. Figure 11\emph{a} shows the explanation of
Type I propagation. The X-point of the magnetic reconnection site
occurs over the middle part of a set of loops (marked by
\emph{Reconnection}), then continuing magnetic reconnection occurs
bi-directionally, thus results in the brightening of the PFLs
appearing in the middle of the set of loops and propagating
towards two opposite directions (see the grey thick arrows) as
expected from the standard model. The white hollow arrows
represent the separating direction of the FRs. Figure 11\emph{b}
illustrates that the magnetic reconnection site occurs over one
end of a set of loops (mark by \emph{Reconnection}), then the
magnetic reconnection develops continuously towards the other end
(see the black thick arrow). This mechanism can be employed to
explain the observations of Type II propagation that the
brightening of the PFLs occurs at one end of a set of PFLs and
propagates from one end to the other one. Type III propagation
displays a complex evolution pattern of the PFLs. We suggest that
there are two (or more than two) magnetic reconnection sites (see
Fig. 11\emph{c}) which locate in two (or more than two) magnetic
flux system. After the magnetic reconnection occurs, each of the
brightening of the PFLs propagates bi-directionally along the
neutral lines (see the black thick arrows). We check the MDI
magnetograms and the TRACE WL observations of these events (e.g.
see the examples displayed in Figs. 3-5), and find that the
magnetic activity are clearly seen at the footpoints of the
initial brightening of the PFLs. According to the observations of
these three types of propagation, we propose that all the
propagation of the PFLs are caused by magnetic reconnection. There
is only one set of magnetic flux system involved in Type I and
Type II propagation. The only difference of these two types of
propagation is the position of the magnetic reconnection site.
There are two or more than two sets of PFLs heated in the process
of flare in Type III propagation. So the different appearance of
these three types of the propagating PFLs may be determined by two
effects. One is the different position of the magnetic
reconnection site, the other is the different magnetic
configuration of the source region of the flare.

Further studies using the data observed by Solar Terrestrial
Relations Observatory (STEREO) will give us a more clear
three-dimensional stereoscopic pictures about the evolution of the
PFLs. Vector magnetograms (e.g. from Hinode) with higher spatial
and temporal resolution are likely to uncover the magnetic
configuration in detail. Moreover, complex three-dimensional
simulation will be required to understand the dynamic behavior of
the PFLs and the FRs.

\acknowledgements

The authors are indebted to the TRACE and SOHO/MDI teams for
providing the data. The work is supported by the National Natural
Science Foundations of China (G10703007, 10573025, 10603008,
40674081, and 10733020), the CAS Project KJCX2-YW-T04, and the
National Basic Research Program of China under grant
G2006CB806303.

\clearpage

\input{tab1.tex}

\begin{deluxetable}{rrrrrrrrr}
\tabletypesize{\scriptsize} \tablecolumns{9} \tablewidth{0pc}
\tablecaption{The 8 parameters and the corresponding flare events}
 \tablehead{ \colhead{~} & \colhead{FC} &
\colhead{FD} & \colhead{PD} & \colhead{Vp} & \colhead{ap}  &
\colhead{Len} & \colhead{SD} & \colhead{Vs}}
\startdata
FC & 190 & 190 & 183 & 183 & 183 & 124 & 101 & 101\\
FD & ~   & 190 & 183 & 183 & 183 & 124 & 101 & 101\\
PD & ~   & ~   & 183 & 183 & 183 & 117 & 94 & 94\\
Vp & ~   & ~   & ~   & 183 & 183 & 117 & 94 & 94\\
ap &~    &~    &~    &~    & 183 & -   & -  & - \\
Len & ~  & ~   &~    &~    &~    & 124 & 101 & 101\\
SD & ~   & ~   &~    & ~   &~    &~    & 101 & 101 \\
Vs &~    &~    & ~   &~    & ~   & ~   &~    & 101 \\
\enddata
\tablenotetext{~}{The number on the cross point of two parameters
represents the flare events from which the corresponding two
parameters are determined simultaneously.}
\end{deluxetable}

\begin{deluxetable}{rrrrrrr}
\tabletypesize{\scriptsize} \tablecolumns{7} \tablewidth{0pc}
\tablecaption{The information of the three types of propagation of
the PFLs and the separation of the FRs.}
 \tablehead{ \colhead{Type} & \colhead{M-class} &
\colhead{X-class} & \colhead{Total} & \colhead{Percentage}    &
\colhead{Vp(Vs)[km s$^{-1}$]} & \colhead{Standard Deviation[km
s$^{-1}$]}} \startdata
I & 76(50) & 18(14) & 94(64) & 49.5$\%$(51.6$\%$) & 18.3(8) & 10.7(6.1)\\
II & 41(23) & 16(13) & 57(36) & 30$\%$(29$\%$) & 16.8(6.3) & 7.4(3.7)\\
III & 34(20) & 5(4) & 39(24) & 20.5$\%$(19.4$\%$) & 17.7(6) &
12.7(5)
\enddata
\tablenotetext{~}{The values in brackets, determined from TRACE
1600 \AA~observations, are relevant to the separation of the FRs.
Otherwise the values are relevant to the propagation of the PFLs,
derived from TRACE 171 \AA~observations.}
\end{deluxetable}

\clearpage

\begin{figure}
\epsscale{1.} \plotone{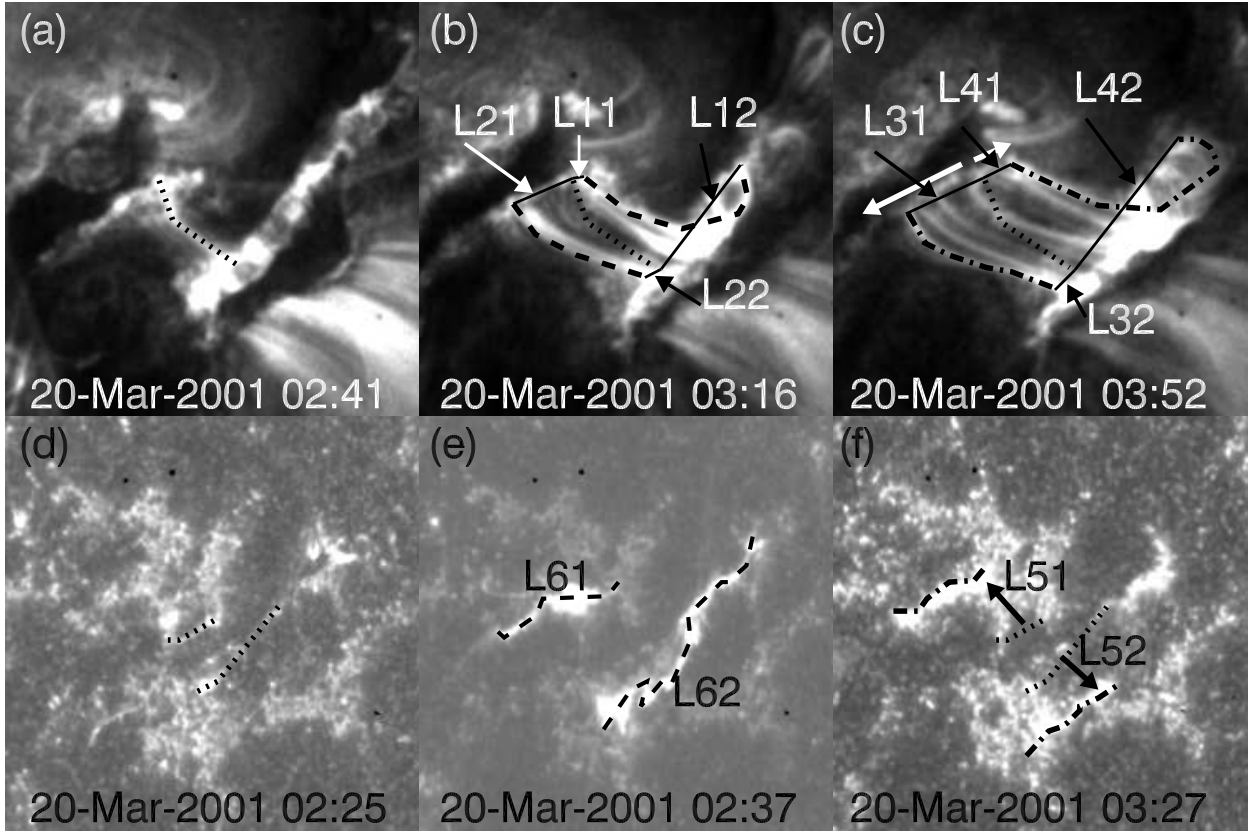} \caption{Time sequence of TRACE 171
\AA~(\emph{a-c}) and 1600 \AA~(\emph{d-f}) images showing a flare
event on 2001 March 20. This event is employed to describe the
method to measure the parameters relevant to the PFLs and the FRs
(see the text). The dotted curves in (\emph{a})-(\emph{c})
represent the initial brightening of the PFLs. The dashed curves
in (b) and (c) outline the outer edges of the PFLs at 03:16 UT,
and the dash dot curves in (\emph{c}), the outer edges at 03:52
UT. \emph{L$_{1}$} and \emph{L$_{2}$} in (\emph{c}) indicate the
distance of the propagation of the PFLs from 02:41 UT to 03:52 UT
towards southeast along the FRs, as well as \emph{L$_{3}$} and
\emph{L$_{4}$}, towards northwest. The thick white arrows in
(\emph{c}) represent the propagating direction of the PFLs. The
dotted curves in (\emph{d}) and (\emph{f}) represent the initial
brightening of the FRs at 02:25 UT. The dash dot curves in
(\emph{f}) outline the outer edges of the FRs at 03:27 UT. The
dashed curves in (\emph{e}) mark the track of the FRs.
\emph{L$_{61}$} and \emph{L$_{62}$} represent the length of the
FRs. \emph{L$_{51}$} and \emph{L$_{52}$} in (\emph{f}) are the
average distance of the separation of the FRs from 02:25 UT to
03:27 UT. The black arrows in (\emph{f}) point to the separating
direction of the FRs. The field-of-view is 125\arcsec $\times$
125\arcsec. \label{f1}}
\end{figure}

\begin{figure}
\epsscale{1.} \plotone{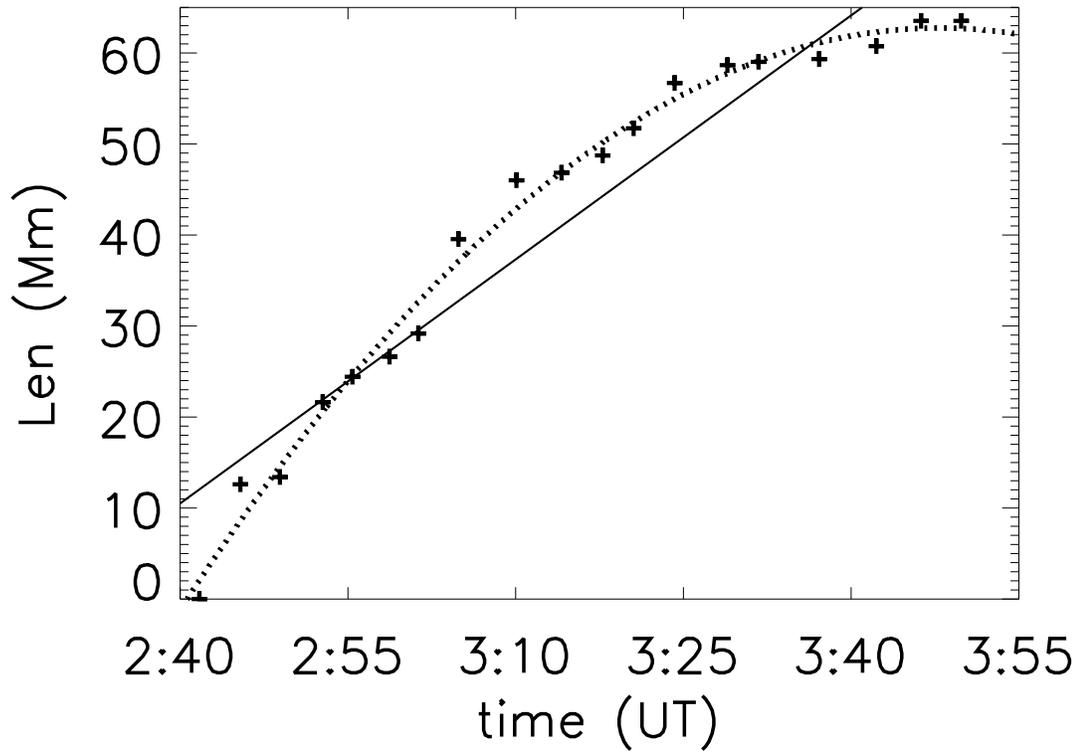} \caption{Distance versus time plot
for the propagation of the brightening of PFLs for the event shown
in Fig. 1. The solid line are linear fit to the data points, and
the dotted line, second order polynomial fit. \label{f2}}
\end{figure}

\begin{figure}
\epsscale{1.} \plotone{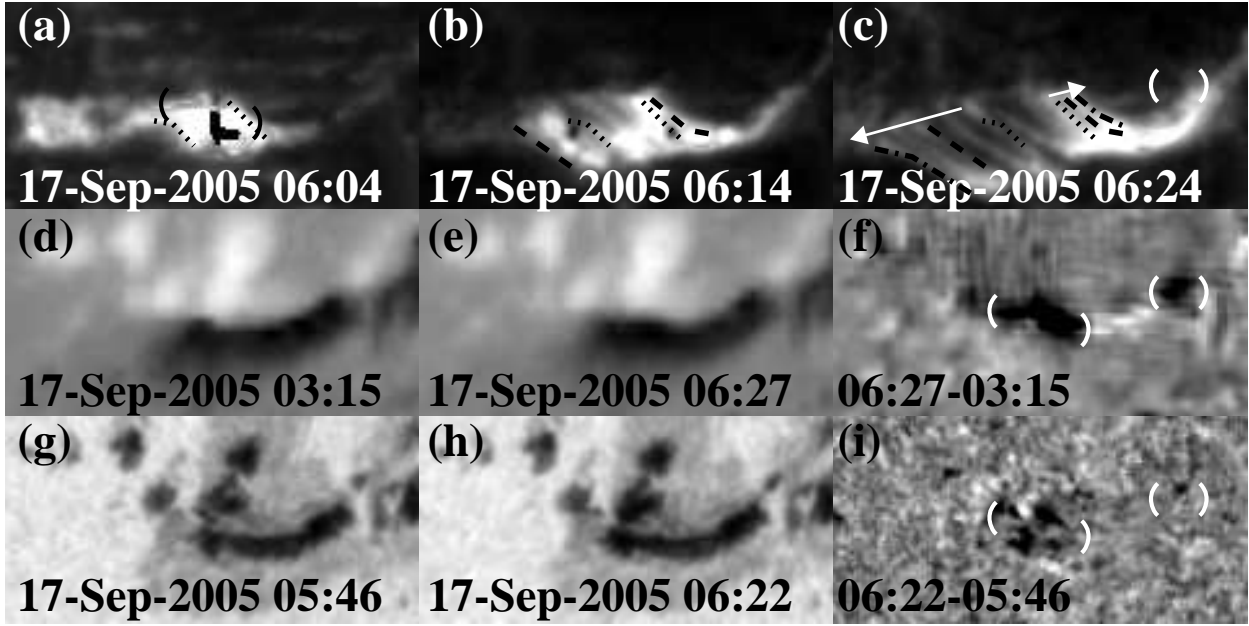} \caption{TRACE and SOHO/MDI
observations showing a flare event on 2005 September 17.
(\emph{a})-(\emph{c}): time sequence of TRACE 171 \AA~images.
(\emph{d})-(\emph{f}): the observations of SOHO/MDI longitudinal
magnetograms (\emph{d} and \emph{e}) and their difference image
(\emph{f}). (\emph{g}) and (\emph{i}): TRACE WL images (\emph{g}
and \emph{h}) and their difference image (\emph{i}). The curves
and arrows represent the same signification as mentioned in Figs.
1a-1c. The brackets in (f) and (i) mark the regions where display
distinct change of photospherical magnetic field and continuum
intensity, and are overplotted on TRACE 171 \AA~images (a) and
(c). The field-of-view is 80\arcsec $\times$ 40\arcsec.\label{f3}}
\end{figure}

\begin{figure}
\epsscale{1.} \plotone{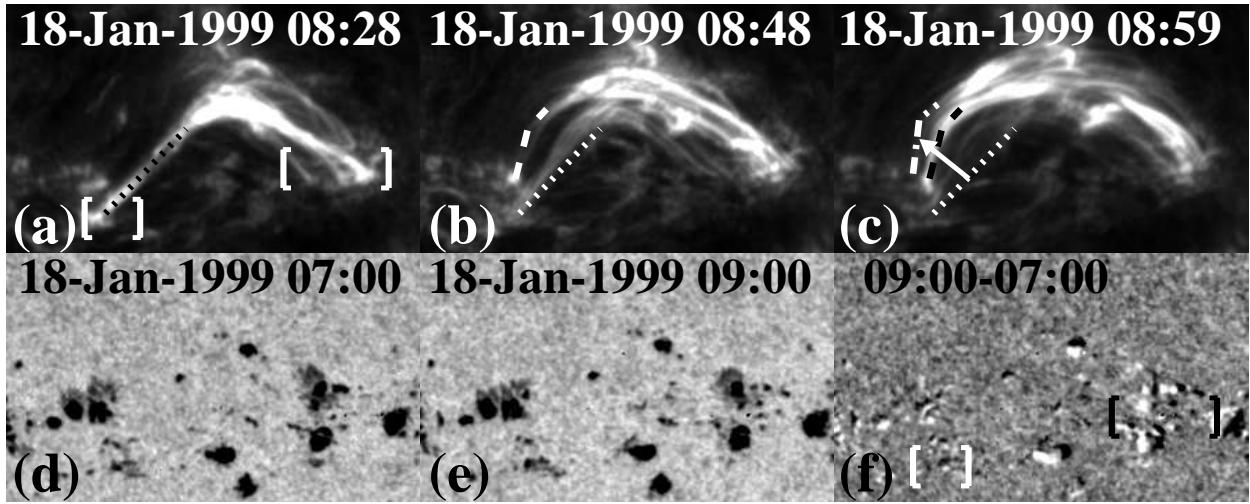} \caption{Similar to Fig. 3. TRACE
observations showing a flare event on 1999 January 18.
(\emph{a})-(\emph{c}): time sequence of 171 \AA~images.
(\emph{d})-(\emph{f}): WL images (\emph{d} and \emph{e}) and their
difference image (\emph{f}). The curves, arrow and brackets denote
the same meaning as in Fig. 2. The field-of-view is 150\arcsec
$\times$ 90\arcsec. \label{f4}}
\end{figure}

\begin{figure}
\epsscale{1.} \plotone{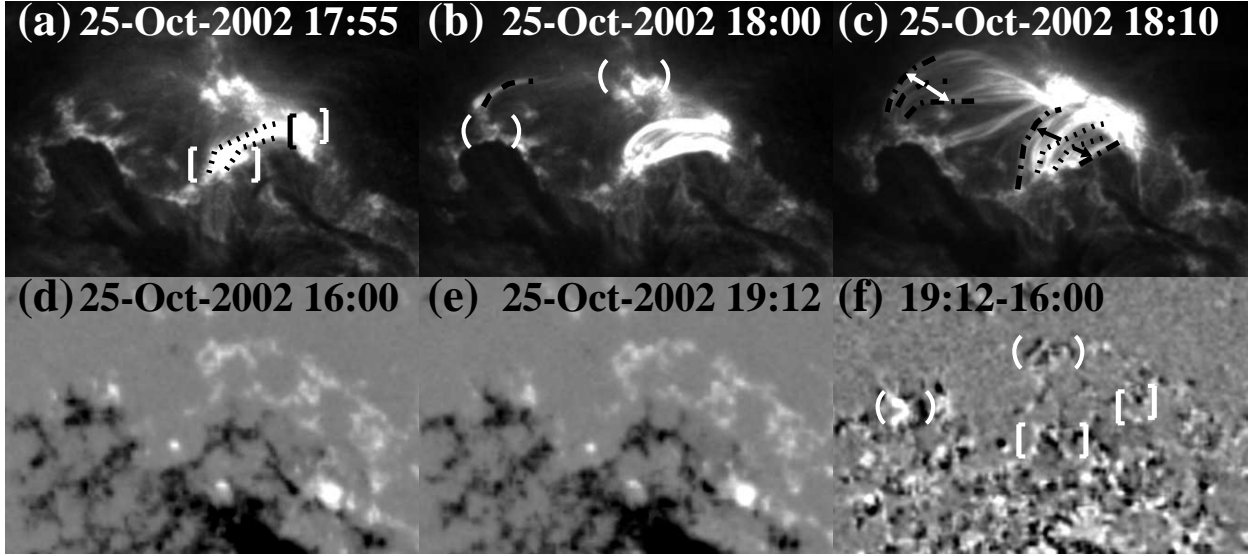} \caption{Similar to Fig. 3. TRACE
and SOHO/MDI observations showing a flare event on 2002 October
25. (\emph{a})-(\emph{c}): time sequence of TRACE 171 \AA~images.
(\emph{d})-(\emph{f}): the observations of SOHO/MDI longitudinal
magnetograms (\emph{d} and \emph{e}) and their difference image
(\emph{f}). The dotted curves in (\emph{a}) represent the outer
edges of the initial brightening of a set of the PFLs in the
southwest of the FRs at 17:55 UT, the dashed curve in (\emph{b}),
outlines the first brightening of another set of the PFLs in the
northeast at 18:00 UT, the dash dot curves in (\emph{c}), the
outer edges of the two set of the PFLs at 18:10 UT. Black and
white arrows in (\emph{c}) represent the propagating direction of
the brightening. The square brackets in (\emph{a}) represent the
region where displays the initial brightening in southwest of the
FRs, and are overplotted on the difference magnetogram (f). The
brackets in (\emph{b}) and (\emph{f}), the region in northeast.
The field-of-view is 225\arcsec $\times$ 150\arcsec.\label{f5}}
\end{figure}

\begin{figure}
\epsscale{1.} \plotone{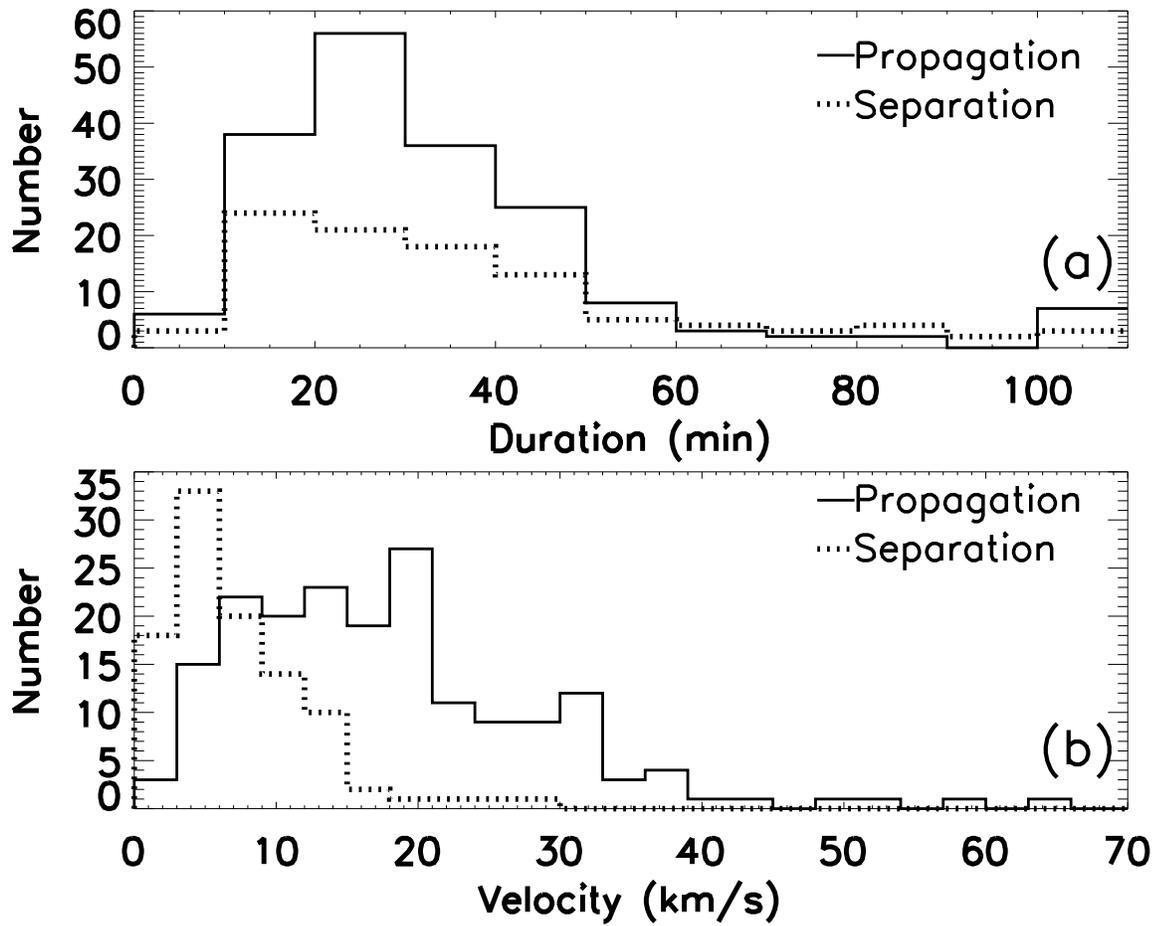} \caption{(\emph{a}): histogram
showing the distribution of \emph{PD} (\emph{solid lines}) and
\emph{SD} (\emph{dotted lines}). (\emph{b}): histogram displaying
the distribution of \emph{V$_{p}$} (\emph{solid lines}) and
\emph{V$_{s}$} (\emph{dotted lines}). \label{f6}}
\end{figure}

\begin{figure}
\epsscale{1.} \plotone{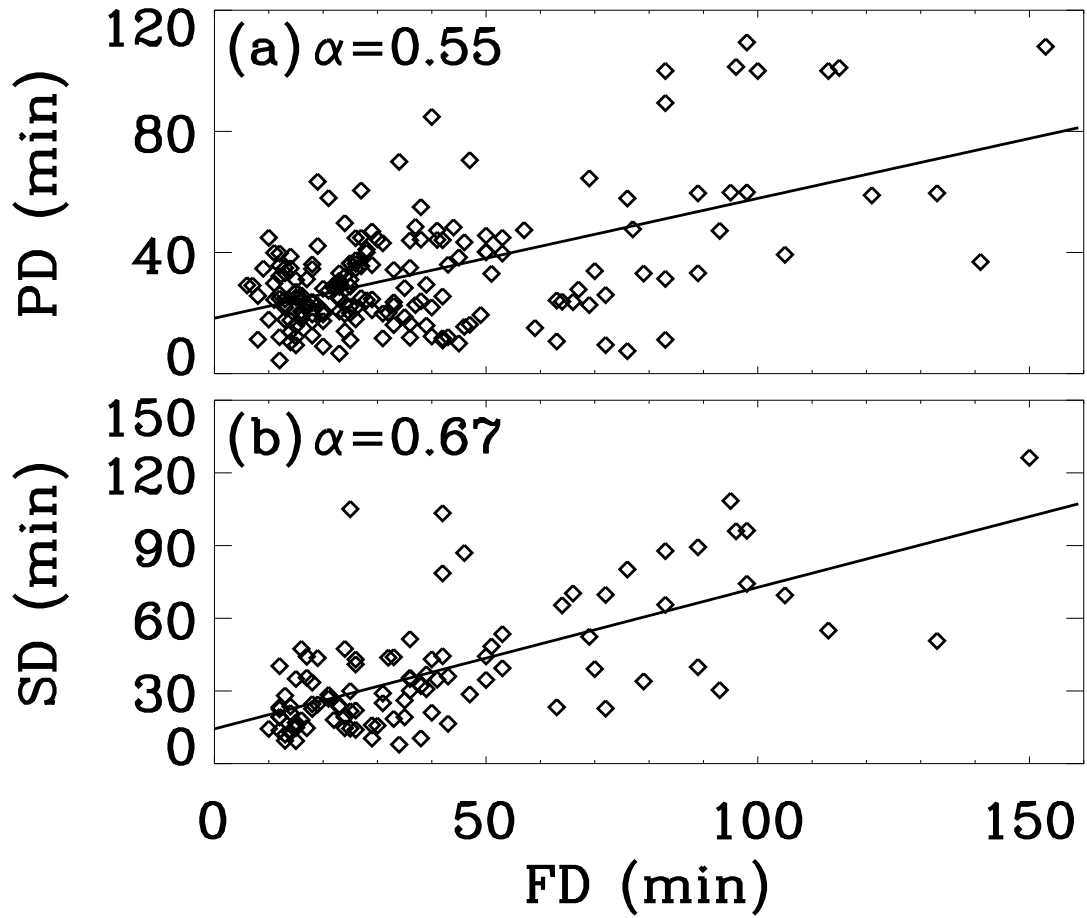} \caption{The relationships between
\emph{PD} and \emph{FD} (\emph{a}), as well as \emph{SD} and
\emph{FD} (\emph{b}). \emph{$\alpha$} represents the correlation
coefficient, and lines show the linear fitting of the data points.
\label{f7}}
\end{figure}

\begin{figure}
\epsscale{1.} \plotone{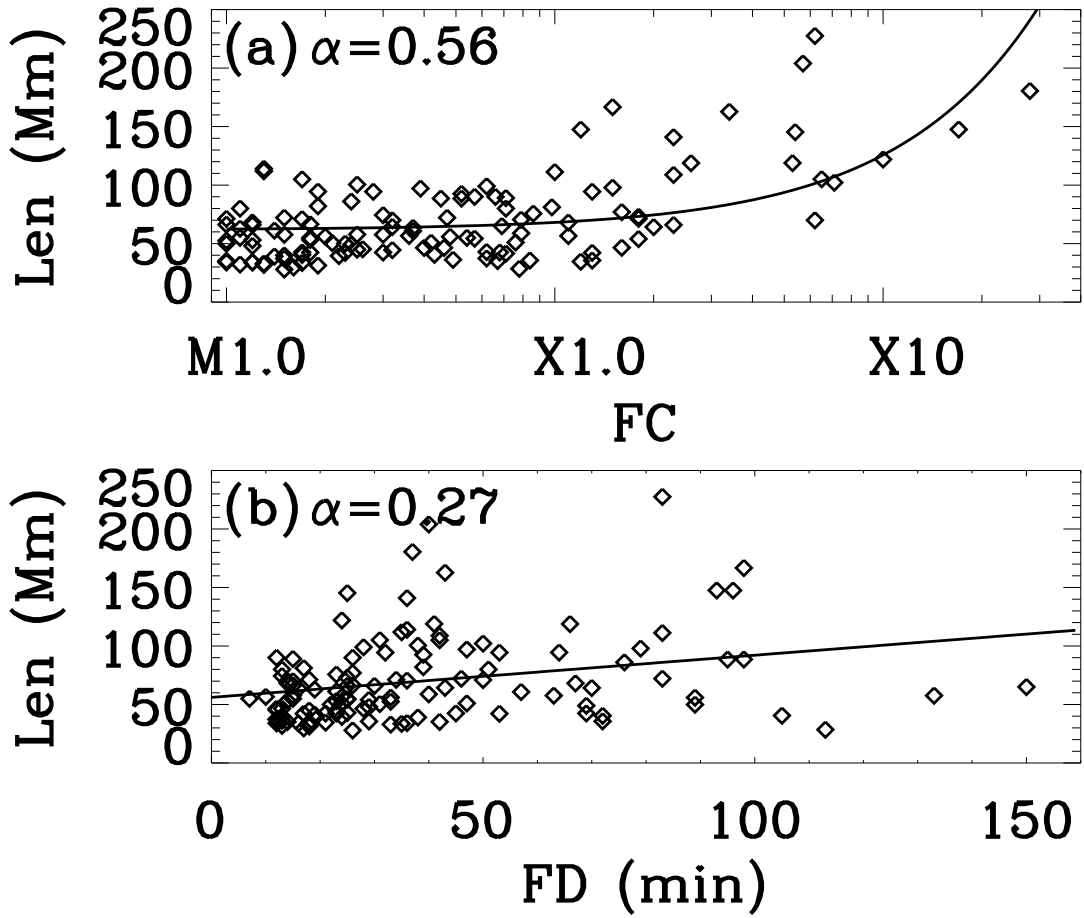} \caption{Same as Fig. 7, but for
the relationships between \emph{Len} and \emph{FC} (\emph{a}), as
well as \emph{Len} and \emph{FD} (\emph{b}). \label{f8}}
\end{figure}

\begin{figure}
\epsscale{1.} \plotone{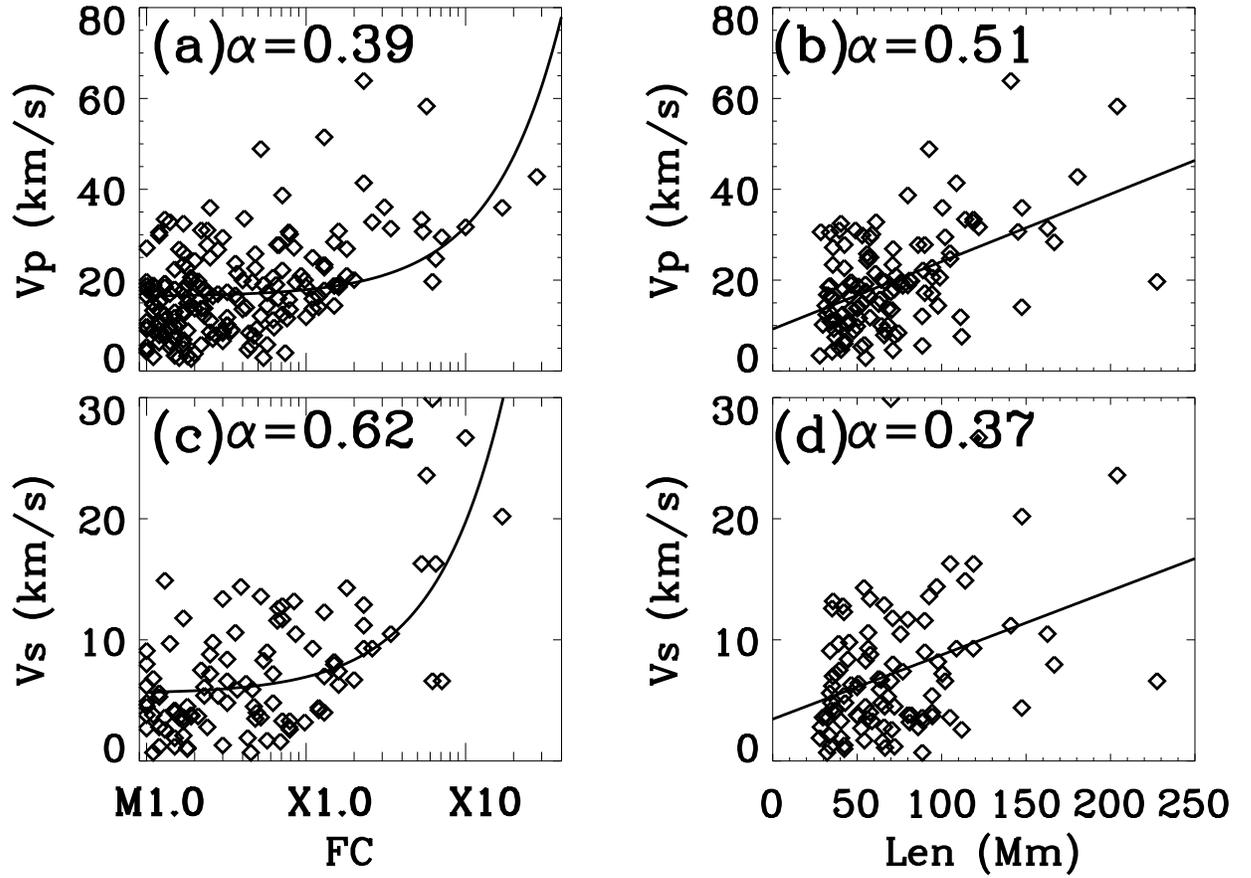} \caption{Same as Fig. 7, but for
the relationships between \emph{V$_{p}$} and \emph{FC} (\emph{a}),
\emph{V$_{p}$} and \emph{Len} (\emph{b}), \emph{V$_{s}$} and
\emph{FC} (\emph{c}), as well as \emph{V$_{s}$} and \emph{Len}
(\emph{d}). \label{f9}}
\end{figure}

\begin{figure}
\epsscale{1.} \plotone{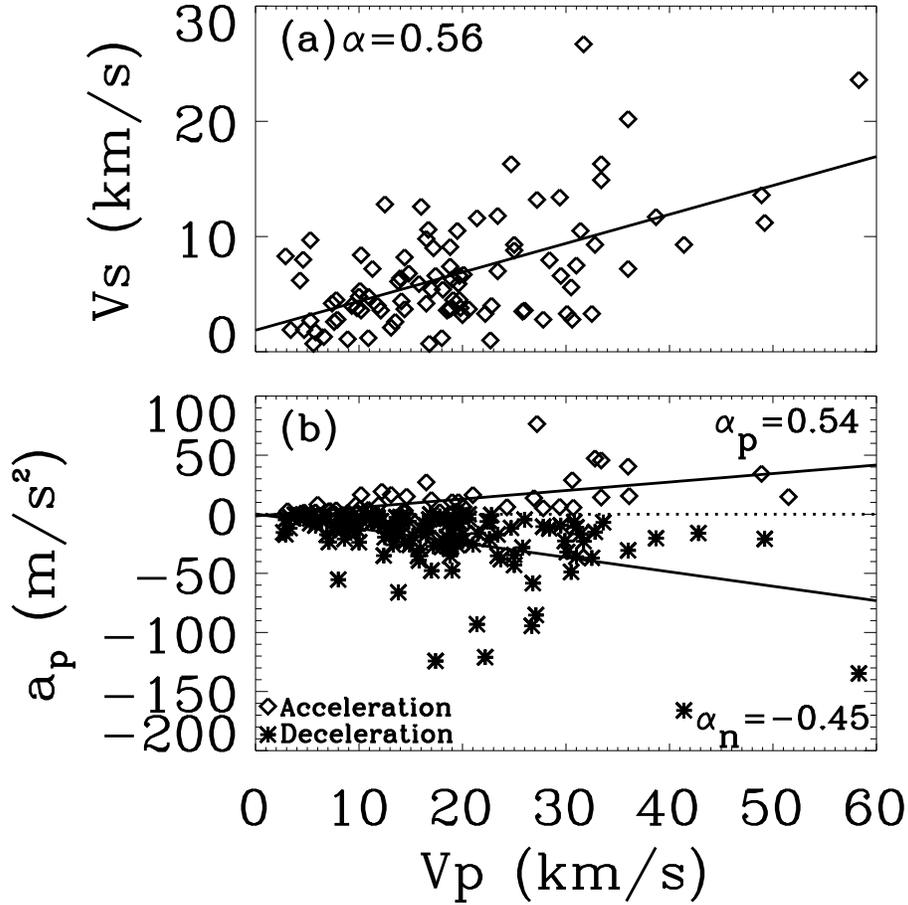} \caption{ Same as Fig. 7, but for
the relationships between \emph{V$_{s}$} and \emph{V$_{p}$}
(\emph{a}), as well as between \emph{a$_{p}$} and \emph{V$_{p}$}
(\emph{b}). The diamonds in (\emph{b}) represent the events with
positive acceleration, and the asterisks with negative
acceleration. \label{f10}}
\end{figure}

\begin{figure}
\epsscale{.7} \plotone{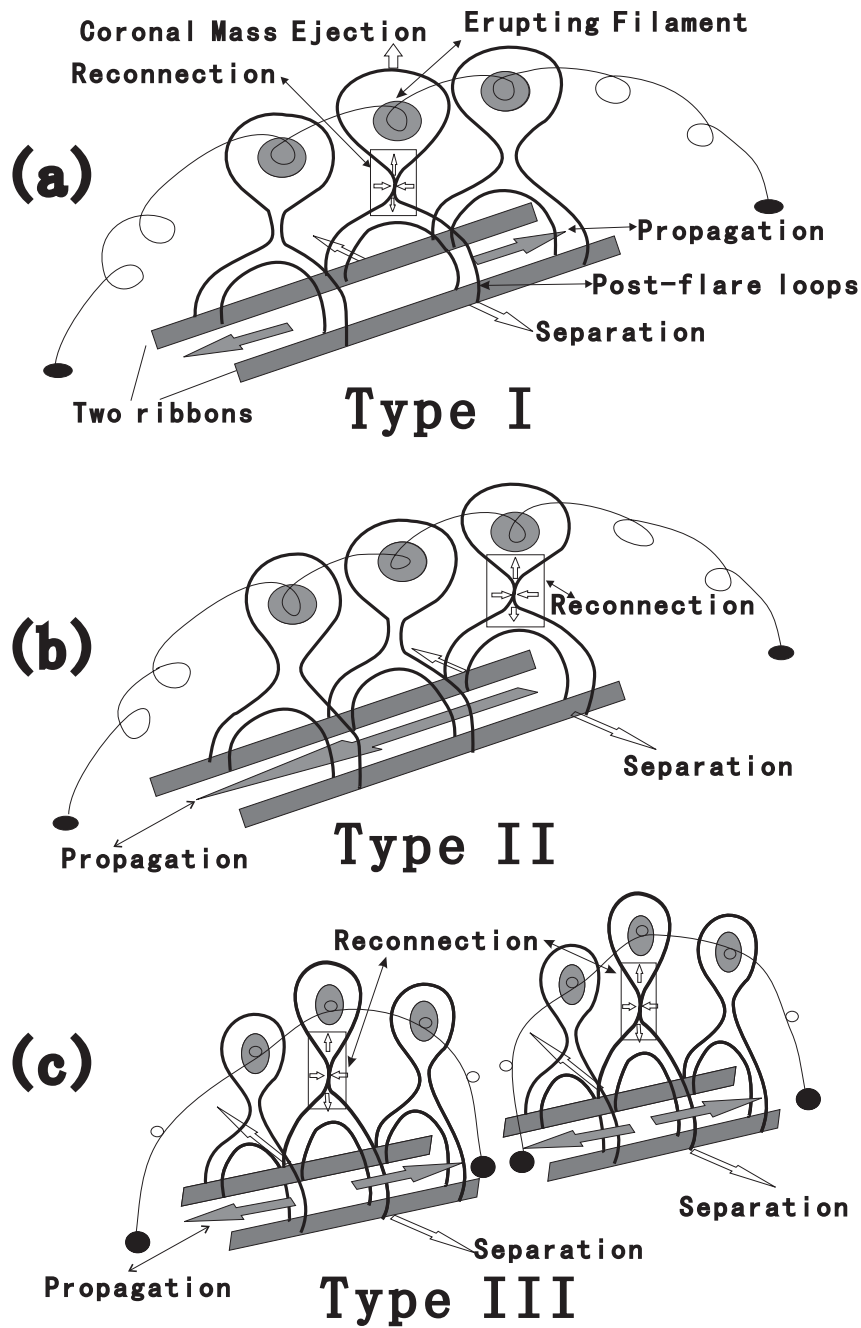} \caption{Schematic diagrams
showing 3-dimensional magnetic reconnection in the process of
flares and erupted filaments. These figures are taken from
\citet{shi05} and \citet{tri06a}, and have been modified according
to TRACE observations. For each figure, the thick curves show
magnetic field lines, and the thin lines, magnetic flux ropes or
filament. The gray thick lines mean the FRs. The region in the
rectangle window represents the magnetic reconnection site. The
arrows in this window means the moving direction of the material.
The bigger grey arrows represent the propagation directions of the
PFLs, and bigger hollow arrows the separating directions of the
FRs.\label{f11}}
\end{figure}

\end{document}

%% file: tab1.tex
\ProvidesFile{table.tex}%
 [2003/12/12 5.2/AAS markup document class]%

\begin{deluxetable}{rrrrrrrrr}
\tabletypesize{\scriptsize} \tablecolumns{9} \tablewidth{0pc}
\tablecaption{The information of the events} \tablehead{
\colhead{Date} & \colhead{PT(UT)}   & \colhead{FC}    &
\colhead{Dur(min)} & \colhead{Len(Mm)}    & \colhead{Vp(km/s)} &
\colhead{ap(m/s$^2$)} & \colhead{Vs(km/s)} & \colhead{Pos}}
\startdata
14-Jul-98& 12:59 & M4.6 & 12 & 46.4 & 19.6 & 10.5 & 5.9 & S23E20 \\
23-Aug-98& 09:34 & M2.2 & 24 & 39.5 & 31 & -11  & 7.5 & N32E33\\
23-Sep-98& 07:13 & M7.1 & 51 & 80 &  38.7 &  -20.3  & 11.7  & N18E09 \\
30-Sep-98& 13:40 & M2.8 & 100 &    \nodata & 26.7 &  -94.4  &  \nodata &   N23W81  \\
05-Nov-98& 19:55 & M8.4 & 72 & 35.7 & 27.2  &  76.5  & 13.2 & N18W21 \\
30-Dec-98& 05:46 & M1.0 & 34  & 71.2 & 4.6,4.5,5.7 & -0.3 & 8 & N28E03   \\
18-Jan-99& 08:04 & M2.0 & 31 &  \nodata & 13.8 & -8.3 & \nodata & N19E03\\
28-Feb-99& 16:39 & M6.6 & 12 & 90 & 21.4 & -93.1 & 11.6 & N28W06\\
10-May-99& 05:31 & M2.5 & 15 & 57.8 & 25 & -43 & 8.8 & N14E02\\
22-Jun-99& 18:29 & M1.7 & 77 & \nodata & 17.8,17 & -1.8 & \nodata & N25E40  \\
02-Jul-99& 01:38 & M2.5 & 18 & 44.9 & 11.1 & -2.8 &\nodata & S27E05  \\
05-Jul-99& 18:48 & M1.2 & 36 & 34.2 & 10.9 & 2.8 & 1.2 & S25W43\\
25-Jul-99& 13:38 & M2.4 & 69 & 49 & 31 & -36 & \nodata & N38W81\\
20-Oct-99& 06:22 & M1.7 & 50 & 71 & 23.4,19.4 & -17.3 & 11.8 & N10W48  \\
26-Oct-99& 18:52 & M2.3 & 22 & 50 & 13.8,12.1 & -66.2 & 6.1 & S13W02  \\
12-Nov-99& 09:16& M1.7 & 42 & 105 & 26 & -4.6 & 3.6 & N10E05 \\
12-Nov-99& 11:54 & M1.1 & 7 & 55 & 17.6 & -3.8 & \nodata & N10E17  \\
16-Nov-99& 14:11 & M1.4 & 49 & \nodata & 6.4,3 & -5.6 & \nodata & S09E06  \\
22-Jan-00& 18:01 & M1.0 & 12 & 34 & 18.8 & -35 & 9.1 & S23W50  \\
08-Feb-00& 09:00 & M1.3 & 36 & 114 & 33.4,13.8 & 45.7 & 14.9 & N25E26 \\
24-Mar-00& 07:52 & X1.8 & 18 & 71.3 & 26.9 & 13.7 & \nodata & N16W82 \\
11-Apr-00& 18:10 & M1.0 & 20 & \nodata & 16.5 & 27.1 & \nodata & S16W18\\
12-Apr-00& 03:35 & M1.3 & 23 &  \nodata & 8.9 & -8.7 & \nodata & S15W24 \\
19-May-00& 00:58 & M1.1 & 13 & 80 & 18.9 & -27.6 & 3.8 & N11E07 \\
04-Jun-00& 14:01 & M3.2 & 14 & 70 & 17.4 & -124.2 & 6.6 & N20E35  \\
06-Jun-00& 14:01 & M7.1 & 15 & 89 & 22.2 & -121 & 3.3 & N19E14 \\
06-Jun-00& 15:25 & X2.3 & 42 & 108.8 & 41.4 & -166 & 9.3 & N19E14 \\
10-Jun-00& 17:02 & M5.2 & 39 & 92.6 & 48.9 & 34.4 & 13.6 & N22W38 \\
23-Jun-00& 04:07 & M2.6 & 12 & 45.3 & 16.5 & -17 & 9.8 & N19W30 \\
25-Jun-00& 07:52 & M1.9 & 64 & 94.5 & 20.6,18.6 & -3.9 & 3.7 & N16W55  \\
10-Jul-00& 20:05 & M1.9 & 18 & 31.3 & 14.4 & 3.1 & 3.7 & N16W43 \\
11-Jul-00& 13:10 & X1.0 & 83 & 111.2 & 11.9 & 7.9 & \nodata & N18E45 \\
12-Jul-00& 05:02 & M1.2 & 14 & 68.2 & 10.1,8.9,4.2 & 4.7 & 5.3 & N16E31  \\
12-Jul-00& 18:49 & M5.7 & 26 & 90.1 & 17.2,2.9 & -27.1 & 9 & N16W64  \\
14-Jul-00& 10:24 & X5.7 & 40 & 204 & 58.3,30.5 & -134.7 & 23.6 & N22W07 \\
14-Jul-00& 13:52 & M3.7 & 16 & 63.5 & 15.8 & -39.1 & 5.9 & N20W08  \\
17-Jul-00& 00:34 & M1.4 & 38 & 39 & 5.3,2.5 & -3.6 & 9.7 & N09E81 \\
18-Jul-00& 05:15 & M1.9 & 39 & 82 & 19.8 & 1.4 & 3.8 & N17W58  \\
25-Jul-00& 18:46 & M1.2 & 16 & 33.9 & 30.5 & -48.8 & 5.6 & N05W16 \\
25-Sep-00& 02:15 & M1.8 & 20 & \nodata & 16.2,11.5 & -7.9 & \nodata & N15W42 \\
21-Oct-00& 18:31 & M3.0 & 33 & \nodata & 8.8,8.8 & -5.1 & \nodata & N17E23  \\
08-Nov-00& 23:28 & M7.4 & 83 & \nodata & 4,8.4,4 & -0.3 & \nodata & N05W75 \\
24-Nov-00& 15:13 & X2.3 & 30 & 66 & \nodata & \nodata & 12.9 & N22W07  \\
24-Nov-00& 21:59 & X1.8 & 29 & 54 & \nodata & \nodata & 14.3 & N21W14  \\
20-Mar-01& 02:18 & M1.1 & 19 & 62.8 & 14.8 & -6.1 & 6.8 & S05W54  \\
20-Mar-01& 15:07 & M1.6 & 17 & 29.7 & 10.2 & 16.5 & 3.6 & S05W61  \\
21-Mar-01& 02:37 & M1.8 & 14 & 53.2 & 14.9 & -25.8 & \nodata & S05W65  \\
24-Mar-01& 19:55 & M1.7 & 72 & 40.4 & 32.5 & -37 & 3.3 & N15E22 \\
28-Mar-01& 12:40 & M4.3 & 105 & 40.5 & 4.7,12.5 & 1.7 & 1.9 & N18E02 \\
29-Mar-01& 14:58 & M1.5 & 13 & 39.9 & 9.9 & -13.6 & \nodata & N16W13\\
31-Mar-01& 11:12 & M2.1 & 31 & 50.5 & 18.7 & 0.1 & 3.7 & N16W34   \\
02-Apr-01& 11:36 & X1.1 & 67 & 68.1 & 13.7 & -14.9 & \nodata & N11W91  \\
03-Apr-01& 23:51 & M1.1 & 47 & \nodata & 14.6 & 15 & \nodata & N16W88   \\
09-Apr-01& 15:34 & M7.9 & 40 & 58.8 & 30.1 & -32 & 3.3 & S21W04 \\
10-Apr-01& 05:26 & X2.3 & 36 & 141 & 49.2 & -20.9 & 11.2 & S23W09  \\
11-Apr-01& 13:26 & M2.3 & 53 & 42.2 & 18.1,10.8 & -21.8 & 5.4 & S22W27  \\
12-Apr-01& 03:04 & M1.3 & 33 & 33 & 18.5 & -3.7 & 3.6 & S22W38   \\
12-Apr-01& 10:28 & X2.0 & 70 & 64.2 & 20.1 & 5.9 & 6.7 & S19W43   \\
19-Apr-01& 11:35 & M2.0 & 33 & 56 & 24.3 & 6.1 & \nodata & N14E60   \\
20-Apr-01& 05:23 & M1.0 & 42 & 35.2 & 4.3,12.1 & -3.8 & 6.2 & N16E63  \\
20-Apr-01& 20:04 & M4.1 & 25 & \nodata & 33.6,19 & -6.9 & \nodata & N15E55   \\
23-Apr-01& 01:28 & M1.0 & 18 & 34.7 & 10.9 & 4 & 4.8 &  N18E18  \\
23-Apr-01& 20:30 & M4.0 & 28 & 46.4 & 13.6 & -2.5 & \nodata & N14E23 \\
24-Apr-01& 05:42 & M2.1 & 14 & \nodata & 19.8 & -7.6 & \nodata & N18E01 \\
24-Apr-01& 22:24 & M1.8 & 16 & \nodata & 3.3 & -11.8 & \nodata & N17E01   \\
25-Apr-01& 13:48 & M2.7 & 20 & \nodata & 8 & -55.3 & \nodata & N18W09   \\
26-Apr-01& 13:12 & M7.8 & 113 & 28.5 & 30.6 & 28.9 & 2.8 & N17W31   \\
27-Apr-01& 19:15 & M1.2 & 29 & 48 & 8.9 & -7 & \nodata & N18W37  \\
05-May-01& 08:56 & M1.0 & 33 & 52.5 & 5.3 & -7.5 & 2.7 & N25W06  \\
12-May-01& 23:35 & M3.0 & 63 & 57.6 & 29.4 & 6.4 & 13.4 & S17E00  \\
13-May-01& 03:04 & M3.6 & 10 & 56.6 & 16.7 & -14.4 & 10.6 & S18W01  \\
15-May-01& 03:00 & M1.0 & 15 & 66.3 & 19.7 & -7.1 & 4.5 & S17W29  \\
04-Jun-01& 22:59 & M1.7 & 35 & 33.6 & 13.1,12.8 & 16.6 & 2.1 & S05W04  \\
15-Jun-01& 10:13 & M6.3 & 19 & \nodata & 9.6 & -5.3 & \nodata & S26E41  \\
19-Jul-01& 10:04 & M1.8 & 25 & 54.9 & 19.1 & -22.7 & 4.5 & S08W62  \\
05-Aug-01& 22:24 & M4.9 & 13 & 36.2 & 9.3,5.7,4.3 & -6.2 & 4 & S20W49  \\
25-Aug-01& 16:45 & X5.3 & 41 & 118.9 & 33.4 & 14.3 & 16.3 & S17E34  \\
30-Aug-01& 17:57 & M1.5 & 26 & 27.9 & 3.4 & -2.7 & 1.9 & S21W28   \\
02-Sep-01& 13:48 & M3.0 & 13 & 74.4 & 8.4 & -12.9 & \nodata & S21W65   \\
15-Sep-01& 11:28 & M1.5 & 50 & \nodata & 13 & -16.8 & \nodata & S21W49   \\
28-Sep-01& 10:14 & M2.4 & 76 & \nodata & 16.9 & -0.2 & \nodata & S18W36   \\
09-Oct-01& 11:13 & M1.4 & 63 & \nodata & 17,27.4 & 12.2 & \nodata & S28E08 \\
19-Oct-01& 01:05 & X1.6 & 26 & \nodata & 30.8 & -6.6 & \nodata & N16W18  \\
19-Oct-01& 02:32 & M1.2 & 26 & \nodata & 15.2 & -18.3 & \nodata & N16W18   \\
19-Oct-01& 16:30 & X1.6 & 30 & \nodata & 18.5 & -9.1 & \nodata & N15W29  \\
23-Oct-01& 02:23 & M6.5 & 23 & \nodata & 19 & -16.5 & \nodata & S18E11  \\
25-Oct-01& 15:02 & X1.3 & 46 & \nodata & 17.8 & 1.9 & \nodata & S16W21  \\
31-Oct-01& 08:09 & M3.2 & 45 & \nodata & 11.8 & -2.9 & \nodata & N11E02   \\
01-Nov-01& 23:52 & M1.1 & 141 & \nodata & 8.9,11.9 & -2.6 & \nodata & N12W23   \\
13-Dec-01& 14:30 & X6.2 & 15 & 70 & \nodata & \nodata & 29.9 & N16E09 \\
22-Feb-02& 00:10 & M4.4 & 38 & \nodata & 8.4 & -0.01 & \nodata & S20W20   \\
14-Mar-02& 01:50 & M5.7 & 24 & \nodata & 14.4,11.1 & 1.1 & \nodata & S12E23   \\
15-Mar-02& 23:10 & M2.2 & 153 & \nodata & 5.8 & 1.8 & \nodata & S08W03 \\
18-Mar-02& 02:31 & M1.0 & 27 & \nodata & 4.4,6.1 & 0.9 & \nodata & S09W47   \\
09-Apr-02& 13:02 & M1.1 & 16 & \nodata & 13.3 & -25.8 & \nodata & N18E32  \\
10-Apr-02& 19:07 & M1.6 & 27 & \nodata & 26.8 & -58.3 & \nodata & N20E31   \\
16-Apr-02& 13:19 & M2.5 & 37 & \nodata & 8.6 & -20.9 & \nodata & N19W79  \\
21-Apr-02& 01:51 & X1.5 & 115 & \nodata & 18.8 & 0.4 & \nodata & S14W84  \\
24-Apr-02& 21:56 & M1.7 & 15 & \nodata & 12.5 & -9.6 & \nodata & N09W49  \\
30-May-02& 17:24 & M1.6 & 28 & \nodata & 2.9 & -4.7 & \nodata & S16E49 \\
01-Jun-02& 03:57 & M1.5 & 11 & \nodata & 7.1 & -23.5 & \nodata & S19E29 \\
02-Jul-02& 20:31 & M1.5 & 22 & \nodata & 7.2,7.3 & 3 & \nodata & S18W46 \\
29-Jul-02& 02:38 & M4.8 & 17 & \nodata & 22.6 & -21 & \nodata & S15W18  \\
29-Jul-02& 10:44 & M4.7 & 46 & 72.1 & 7.8 & 3.5 & 4.5 & S15W18 \\
31-Jul-02& 01:53 & M1.2 & 23 & 53 & 29.9 & -22.6 & \nodata & S13W30 \\
03-Aug-02& 19:07 & X1.0 & 12 & \nodata & 18.5 & -19.3 & \nodata & X16W76 \\
16-Aug-02& 12:32 & M5.2 & 95 & 88.5 & 12.1 & -5.9 & 3.6 & S14E20 \\
16-Aug-02& 22:12 & M1.2 & 8 & \nodata & 6.3 & -5.1 & \nodata & S14E20 \\
16-Aug-02& 23:33 & M1.7 & 6 & \nodata & 4.9 & -1.1 & \nodata & S05E06 \\
17-Aug-02& 01:08 & M1.1 & 16 & \nodata & 3.1 & 2.8 & \nodata & S19E77 \\
17-Aug-02& 20:51 & M3.4 & 18 & \nodata & 9.1 & -13.2 & \nodata & S06W05  \\
18-Aug-02& 10:05 & M2.3 & 15 & \nodata & 12.2 & 19.2 & \nodata & S07W12  \\
18-Aug-02& 14:39 & M1.9 & 10 & \nodata & 2.7 & -15.4 & \nodata & S06W15  \\
18-Aug-02& 21:25 & M2.2 & 25 & \nodata & 13.8 & -11.9 & \nodata & S12W19 \\
20-Aug-02& 02:57 & M1.4 & 8 & \nodata & 9.7 & 7.8 & \nodata & S10W36 \\
22-Aug-02& 01:57 & M5.4 & 18 & \nodata & 19 & 10.4 & \nodata & S07W62   \\
24-Aug-02& 01:12 & X3.1 & 42 & \nodata & 36.1 & 15.5 & \nodata & S02W81  \\
29-Sep-02& 06:39 & M2.6 & 9 & \nodata & 7.1 & 2.8 & \nodata & N12E21 \\
15-Oct-02& 14:22 & M1.0 & 29 & \nodata & 18,14.2 & -27.6 & \nodata & N20W04 \\
22-Oct-02& 15:35 & M1.0 & 23 & \nodata & 27.1 & -85.3 & \nodata & N28E27  \\
25-Oct-02& 17:47 & M1.5 & 59 & \nodata & 22.4,22.2 & -6.1 & \nodata & N28W11  \\
28-Oct-02& 12:05 & M1.7 & 11 & \nodata & 6.9 & -17.3 & \nodata & N23W61   \\
22-Jan-03& 04:44 & M1.2 & 15 & \nodata & 13.3,7.5 & -20.8 & \nodata & N15W05  \\
19-Mar-03& 03:07 & M1.5 & 83 & 72.1 & 18 & -25.4 & 1.2 & S16W66  \\
27-May-03& 23:07 & X1.3 & 17 & 42.2 & \nodata & \nodata & 12.3 & S07W17  \\
29-May-03& 01:05 & X1.2 & 21 & 34.6 & 16.5 & -9.6 & 4.2 & S06W37  \\
31-May-03& 02:24 & M9.3 & 27 & \nodata & 21 & -25.6 & \nodata & S07W65  \\
06-Jun-03& 23:38 & M1.0 & 27 & \nodata & 9.2 & -5.6 & \nodata & N13E17  \\
11-Jun-03& 00:02 & X1.3 & 53 & 94.4 & 22.8,7.8 & -0.6 & 4 & N10W40   \\
11-Jun-03& 20:14 & X1.6 & 26 & 77 & 18.8 & -29 & 7.4 & N14W57   \\
17-Jun-03& 22:55 & M6.8 & 45 & 42.5 & 27.8 & -10.5 & \nodata & S07E55  \\
10-Jul-03& 14:12 & M3.6 & 29 & \nodata & 23.7 & -38 & \nodata & N12W82  \\
16-Sep-03& 22:24 & M1.3 & 121 & \nodata & 11.2 & -0.6 & \nodata & S12W83   \\
22-Oct-03& 08:44 & M1.7 & 23 & 42.5 & 6 & 8.2 & \nodata & N07E25   \\
23-Oct-03& 20:04 & X1.1 & 24 & 56.6 & 25 & -34.4 & 9.3 & S17E84  \\
24-Oct-03& 02:54 & M7.6 & 47 & 51 & 11.6 & 0.1 & \nodata & S19E72 \\
26-Oct-03& 06:54 & X1.2 & 96 & 147.5 & 14.1 & -5 & 4.4 & S15E43   \\
28-Oct-03& 11:10 & X17 & 93 & 147.6 & 36 & -30.6 & 20.2 & S16E26  \\
29-Oct-03& 20:49 & X10 & 24 & 122 & 31.7 & -22.9 & 26.7 & S15W02   \\
01-Dec-03& 22:38 & M3.2 & 23 & 44.5 & 10.2 & -4.7 & 8.4 & S12W60   \\
04-Dec-03& 19:50 & X28 & 37 & 180.5 & 42.8 & -16 & \nodata & S19W83  \\
17-Dec-03& 09:05 & M4.2 & 24 & 50.8 & 14 & -2.4 & 6.4 & S01E33   \\
18-Dec-03& 07:52 & M3.2 & 43 & 64.3 & 10 & -23.8 & 4.8 & N00E18   \\
18-Dec-03& 08:31 & M3.9 & 47 & 97.1 & \nodata & \nodata & 14.4 & N00E18  \\
18-Dec-03& 10:11 & M4.5 & 98 & 88.6 & 5.6 & -1.1 & 0.7 & N00E08   \\
05-Jan-04& 03:45 & M6.9 & 150 & 65 & \nodata & \nodata & 1.6 & S10E24  \\
26-Feb-04& 02:03 & X1.1 & 20 & \nodata & 15.9 & 14.2 & -29.4 & N14W15  \\
26-Feb-04& 22:30 & M5.7 & 25 & 54.3 & 5.8 & -9.5 & 1.7 & N14W26   \\
06-Apr-04& 13:28 & M2.4 & 76 & 86.1 & 27.8 & 5.2 & 2.8 & S18E15  \\
13-Jul-04& 00:17 & M6.7 & 14 & 35.2 & 16 & -12.4 & 12.6 & N15W49   \\
13-Jul-04& 08:48 & M5.4 & 15 & 55 & 2.9 & -17.1 & 8.3 & N12W52 \\
13-Jul-04& 19:32 & M6.2 & 12 & 37.4 & 11.3 & -6.7 & 7.2 & N13W56 \\
14-Jul-04& 05:23 & M6.2 & 25 & 42.9 & \nodata & \nodata & 4.8 & N12W62  \\
15-Jul-04& 18:24 & X1.6 & 13 & 46.5 & 19 & -47.6 & 6.3 & S11E56   \\
16-Jul-04& 02:06 & X1.3 & 29 & 35.6 & 23.4 & -35.2 & 7 & S10E36 \\
18-Jul-04& 02:57 & M1.5 & 12 & 37.2 & 7.4 & 0.8 & 4.2 & S12E12   \\
20-Jul-04& 12:32 & M8.6 & 23 & 75.7 & 19.5 & -26.3 & 10.5 & N11E34   \\
25-Jul-04& 05:51 & M7.1 & 19 & 41.5 & 12.5 & -19.4 & 12.8 & N10W31   \\
27-Jul-04& 00:00 & M1.2 & 25 & 66 & 7.9,7.4 & -10.9 & 2.8 & N10W54   \\
18-Aug-04& 17:40 & X1.8 & 25 & 73 & 21 & 16.4 & \nodata & S12W83   \\
31-Aug-04& 05:38 & M1.4 & 23 & 61.2 & 32.8 & 47.3 & \nodata & N04W87   \\
12-Sep-04& 00:56 & M4.8 & 89 & 55.9 & 25.8,13.1 & -28.2 & 3.5 & N04E42   \\
14-Sep-04& 09:30 & M1.5 & 133 & 57.6 & 11.8,6.4 & -5 & 4 & N04E17   \\
15-Jan-05& 23:02 & X2.6 & 66 & 118.8 & 32.8 & -15 & 9.3 & N14W09   \\
20-Jan-05& 07:01 & X7.1 & 50 & 102.1 & 29.5 & -9.5 & 6.6 & N12W58   \\
06-May-05& 11:28 & M1.3 & 24 & \nodata & 19.3 & 2.5 & \nodata & S04W76   \\
17-May-05& 02:39 & M1.8 & 21 & 42.4 & 22.7,7.2 & -6.4 & 1 & S15W00   \\
03-Jun-05& 04:11 & M1.3 & 13 & 31.9 & 12.4 & -35 & \nodata & S17E22   \\
16-Jun-05& 20:22 & M4.0 & 41 & \nodata & 20 & -17 & \nodata & N08W90  \\
09-Jul-05& 22:06 & M2.8 & 32 & 94.4 & 17 & -47.7 & 5.4 & N12W28 \\
27-Jul-05& 05:02 & M3.7 & 57 & 60.8 & 21.5 & -16.2 & \nodata & N11E90  \\
30-Jul-05& 06:35 & X1.3 & 44 & \nodata & 51.5,39.5 & 14.8 & \nodata & N12E60   \\
31-Jul-05& 12:24 & M1.1 & 18 & 32.2 & 16.8 & -4.4 & 0.7 & N13E45   \\
01-Aug-05& 13:51 & M1.0 & 89 & 49.9 & 9.8 & 1.7 & 3.8 & N13E32  \\
08-Sep-05& 21:06 & X5.4 & 25 & 145.3 & 30.7 & 5.7 & \nodata & S12E75   \\
09-Sep-05& 05:03 & M1.8 & 26 & 66.5 & 8.9 & 1.2 & 1.1 & S12E68   \\
09-Sep-05& 05:48 & M6.2 & 28 & 99 & 20.6 & -0.1 & \nodata & S23E71  \\
09-Sep-05& 20:04 & X6.2 & 83 & 227.6 & 19.7 & -1.1 & 6.6 & S12E67   \\
11-Sep-05& 02:35 & M3.4 & 11 & \nodata & 8.9 & -3.8 & \nodata & S12E41   \\
11-Sep-05& 13:12 & M3.0 & 69 & 42.5 & 6.6 & -1 & 1.3 & S13E41  \\
13-Sep-05& 19:27 & X1.5 & 98 & 166.7 & 28.4 & -12 & 8 & S11E17   \\
16-Sep-05& 17:48 & M1.3 & 35 & 111.9 & 7.6 & -10.6 & 2.6 & S11W36  \\
17-Sep-05& 06:05 & M9.8 & 17 & 81.1 & 20 & -22.7 & 3.2 & S11W51   \\
02-Dec-05& 10:12 & M7.8 & 20 & \nodata & 15.7 & -35.3 & \nodata & S04E13   \\
02-Dec-05& 21:19 & M2.0 & 40 & \nodata & 3.8 & -0.5 & \nodata & S02E08  \\
27-Apr-06& 15:52 & M7.9 & 36 & 70.5 & 13.5,11 & -12.9 & 2.6 & S08E26   \\
06-Jul-06& 08:36 & M2.5 & 38 & 100.5 & 36 & 40.5 & 7.2 & S09W34   \\
06-Dec-06& 18:47 & X6.5 & 31 & 105.1 & 24.7 & -11.6 & 16.3 & S05E64  \\
13-Dec-06& 02:40 & X3.4 & 43 & 162.7 & 31.4 & -16.6 & 10.5 & S06W23 \\
14-Dec-06& 22:15 & X1.5 & 79 & 97.9 & 14.4 & -23.4 & 8.2 & S06W46\\
\enddata
\end{deluxetable}
